\documentclass[letterpaper, 10 pt, conference]{ieeeconf}  

\IEEEoverridecommandlockouts                              
\usepackage{graphicx} 
\usepackage{multirow}
\usepackage{amsmath}
\usepackage{amsfonts}
\usepackage{float}

\newtheorem{thm}{Theorem}

\newtheorem{assump}{Assumption}
\newtheorem{rem}{Remark}

\title{\LARGE \bf
Distributed consensus on minimum time rendezvous via cyclic
alternating projection }

\author{Hu Chunhe and Chen Zongji
\thanks{This work is partially
supported by the National Natural Science Foundation of China Grant
\#61273349 to Zhou Rui and Grant \#61175109 to Wu Jiang}
\thanks{Hu Chunhe and Chen Zongji are with School of Automation Science and Electrical Engineering, Beihang University(BUAA), Beijing, 100191,
China. The authors are also affiliated with National Key Laboratory
of Science and Technology on Aircraft Control, BUAA.
        {\tt\small Email:hchh@asee.buaa.edu.cn,czj@buaa.edu.cn.}}%
}

\begin{document}

\maketitle
\thispagestyle{empty}
\pagestyle{empty}

\begin{abstract}

In this paper, we propose a distributed algorithm to solve planar
minimum time multi-vehicle rendezvous problem with non-identical
velocity constraints on cyclic digraph (topology). Motivated by the
cyclic alternating projection method that can compute a point's
projection on the intersection of some convex sets, we transform the
minimum time rendezvous problem into finding the distance between
the position plane and the intersection of several second-order
cones in position-time space. The distance can be achieved by metric
projecting onto the plane and the intersection persistently from any
initial point, where the projection onto the intersection is
obtained by Dykstra's alternating projection algorithm. It is shown
that during the procedure, vehicles use only the information from
neighbors and can apply the projection onto the plane
asynchronously. Demonstrations are worked out to illustrate the
effectiveness of the proposed algorithm.
\end{abstract}

\section{INTRODUCTION}

In the past decades, a large amount of attention has been devoted to
coordinate control of multi-vehicle
systems\cite{cao2012overview,ren2005survey,kranakis2006mobile}. The
main object of coordinated control is to allow the multi-vehicle
work together and coordinate their behaviors in a cooperative
fashion to achieve a common goal efficiently. Multi-vehicle
coordination control consists of widespread research fields,
including mission assignment, formation control, rendezvous control,
consensus and distributed estimation, etc. As a fundamental problem,
rendezvous control has attracted great deal of attention from
numerous researchers
\cite{mclain2001cooperative,lin2003multi,lin2004multi,cortes2006robust,fang2008multi,hui2011finite}.
Roughly speaking, it gives the method to drives the vehicles to the
same location.

To achieve rendezvous with time-optimal cost becomes attractive. The
time-optimal rendezvous was firstly introduced as the n-dimension
phase space rendezvous problem for linear systems in
\cite{meschler1963time,chyung1973time}, where the authors concluded
that the intersection of convex attainable collections is the
optimal rendezvous point. In the last decade, the majority of
minimum time rendezvous problems were concentrated on physical
entities (wheeled vehicles \cite{brown2011determination}, unmanned
aerial vehicles (UAVs) \cite{bhatia2008decentralized}, under water
vehicles \cite{crispin2008interception}, spacecraft
\cite{luo2006optimization}, etc.). Naturally, the minimum time
rendezvous can be formed into a standard optimal problem with
dynamics as constraints and time as cost function, which has been
solved by the maximum principle of Pontryagin
\cite{paiewonsky1966three}, dynamic optimization
\cite{burns2007autonomous}, and nonlinear programming
\cite{crispin2008interception}. However, those algorithms are
difficult to solve, especially the two-point value boundary problem,
and rely on off-line centric calculation.

The searching algorithm via level set methods in
\cite{brown2011determination} quantizes the entire 2D environment
with arrival time for each vehicle, and then a min-max method is
applied for the minimum time rendezvous point. A direct heuristic
search algorithm based on path planning is proposed in
\cite{wilson2013real}, where the author analyzes two Dubins vehicles
leader-follower configuration. Searching methods provide a simple
description to the problem but require the knowledge of all
potential rendezvous points.

Most of the literatures mentioned above considered just two
vehicles, moreover, with one on a fixed known trajectory. The
minimum time rendezvous for multi-vehicle, namely more than three
vehicles, brings in new topics and becomes a challenge in the
distributed setting. The decentralized algorithm for Dubins vehicles
to a fixed rendezvous point with the arrival angle as optimization
variable is investigated in \cite{bhatia2008decentralized}. The
distributed consensus algorithm for identical speed multi-agent
time-optimal rendezvous has been studied in
\cite{notarstefano2006distributed}, where the centers for the
minimal enclosing ball and minimal enclosing orthotropic are chosen
as the rendezvous point. Through distributed computation of the
minimal en-closing shapes, consensus control approach provides an
efficient solution.

This paper proposes a novel algorithm for minimum time rendezvous
problem with non-identical velocity constraints for multi-vehicle in
2-D space. Velocity constraint is an essential character for
vehicles, especially for UAVs. When different velocity constraints
applied to different vehicles, the conclusion obtained in
\cite{notarstefano2006distributed} is no longer applicable here.

Studying the minimum time rendezvous problem we show that, any
vehicle with velocity constraint has a bounded reachable distance in
limited time which forms a second-order cone in position-time space.
Furthermore, the minimum time can be acquired by finding the
distance between the position plane and the intersection of those
second-order cones belonging to the vehicles.

Our main result is the design of an algorithm based on the
alternating projection method for the distributed computation on
minimum time rendezvous point. Alternating algorithm is widely
applied to optimal approximation, e.g.,solving linear system
\cite{tam2012method}, linear programming \cite{tam2012method},
signal processing \cite{combettes2011proximal} and even Sudoku
puzzle \cite{tam2012method}.In our algorithm, we utilize Bregman's
alternating projection to obtain the distance represent the minimum
time aforementioned. During the procedure, the metric projection
onto the intersection is necessary, so we employ another projection
algorithm$-$Dykstra's alternating projection as the intermediate
procedure. We show that in our algorithm, vehicles can have the
consensus on the minimum time rendezvous point with only the
information from neighbors on a cyclic interaction topology.
Although only the 2-D space assumption has made in this paper, the
algorithm can be easily extended to higher dimensions rendezvous
problem.

This paper is organized as follows. In Section II, we introduce the
minimum time rendezvous problem and the methods of alternating
projection. Section III presents the geometric description on the
problem and proposes the distributed algorithm on minimum time
rendezvous. Demonstrations are provided as the proof of algorithm's
efficiency in Section IV. Finally in Section V, conclusions are
provided.

\section{Problem Formulation and Preliminary}
\subsection{Minimum time rendezvous with velocity constraints}
In this section, we will introduce the problem of planar minimal
time rendezvous with different velocity constraints. Multi-vehicle
rendezvous problem focuses on the task that how the vehicles can
come together in centralized or decentralized manners. In
\cite{notarstefano2006distributed}, the authors introduced a
identical speed minimum time rendezvous problem, which can be
explained as following:

\begin{equation}
\label{MTRPhomovel} \mathop {\min }\limits_{x,y} \left\{ {\mathop
{\max }\limits_i \left\| {\left( {x,y} \right) - \left(
{{x_i},{y_i}} \right)} \right\|} \right\},
\end{equation}
where $\left( x,y \right)$ , $\left( {{x}_{i}},{{y}_{i}}
\right)~~i\in 1,2,3,\ldots,N $ represent the position of some point
$p$ and vehicle $i$ respectively and $\left\| \cdot  \right\|$ is
the norm. With the homogenous velocity assumption, the optimal
solution consists of moving toward the center of the minimal
enclosing ball(bound on ${{L}_{2}}$ norm) or toward the center of
the minimal enclosing orthotope (bound on the infinity norm) of the
points located at the initial position of the
vehicles.\textsuperscript{\cite{notarstefano2006distributed}}

However, if vehicles have non-identical velocities, the points
mentioned before are no longer the proper minimum time rendezvous
points. This can be attributed to following simple example. Consider
two vehicles with initial points $\left( {{x}_{1}},{{y}_{1}}
\right)$ and $\left( {{x}_{2}},{{y}_{2}} \right)$ in a 2-D Euclidean
space. If two vehicles are at equal velocities, the middle point
$\left( \frac{{{x}_{1}}+{{x}_{2}}}{2},\frac{{{y}_{1}}+{{y}_{2}}}{2}
\right)$ between these two vehicles is the minimum time rendezvous
point, which is indeed the center of the minimal enclosing ball or
the minimal enclosing orthotope. If the vehicles have different
velocities ${{v}_{1}}$, ${{v}_{2}}$, the minimum time rendezvous
point becomes $\left(
\frac{{{v}_{1}}{{x}_{2}}+{{v}_{2}}{{x}_{1}}}{{{v}_{1}}+{{v}_{2}}},\frac{{{v}_{1}}{{y}_{2}}+{{v}_{2}}{{y}_{1}}}{{{v}_{1}}+{{v}_{2}}}
\right)$.

Therefore, we have to transform the minimum time rendezvous problem
for multi-vehicle into

\begin{equation}
\label{MTRPhetervel} \mathop {\min }\limits_{x,y} \left\{ {\mathop
{\max }\limits_i \frac{{\left\| {\left( {x,y} \right) - \left(
{{x_i},{y_i}} \right)} \right\|}}{{{v_i}}}} \right\},
\end{equation}
where $\left( {{x}_{i}},{{y}_{i}} \right)~~i\in 1,2,3,\ldots,N $ and
$v_i$ are the initial point and the velocity of vehicle $i\in
1,2,3,\ldots,N $ respectively and $\left( x,y \right)$ is an
arbitrary point.

We make further assumptions to this min-max problem in distributed
setting:

\begin{assump}
The vehicles can not start from the same position and each vehicle
$i$ has fixed velocity ${{v}_{i}}>0$, $i\in 1,2,3,\ldots,N $.
\end{assump}

\begin{assump}
The vehicles are memoryless except for the initial position and can
only access their own state including position and velocity without
interaction.
\end{assump}

\begin{assump}
The communication between the vehicles is limited in a cyclic
digraph interaction topology. Without loss of generality, the
interaction sequence is according to the number assignment to the
vehicles, i.e., vehicle $i$ receiving information from $i-1$ ,$i\in
2,3,\ldots,N$ and vehicle $1$ receiving from $N$.
\end{assump}

Under the assumptions, each vehicle receives the estimate to the
rendezvous point from the previous one, executes calculation and
sends to the next.

\subsection{Convex sets intersection seeking method via Alternating Projection}

Alternating projection algorithm is a type of geometric optimization
method. Through iteratively orthogonally projecting onto finite many
Hilbert spaces successively in cyclic setting, the limit to the
projection sequence provides an approximation of the initial point
to those spaces.

Bregman's alternating projection, known as Bregman's algorithm or
Bregman's method designed for closed convex sets is always used to
obtain a point in the intersection of convex sets. In considering
two convex sets without intersection, Bregman's algorithm achieve
the distance between the two sets \cite{boyd2003alternating}.
Following theorem provides detailed descriptions.

Assume there are two convex sets $A,B\subseteq{\mathbb{R}^{n}}$, and
${{P}_{A}}\left( \cdot  \right),{{P}_{B}}\left( \cdot \right)$
denote projection on $A$ and $B$, respectively. We have the
following theorem on above sequences:

\begin{thm}\textsuperscript{\cite{cao2012overview}}
Let $A,B\subseteq {{R}^{n}}$ be closed convex sets and $\left\{
{{a}_{n}} \right\}_{n=1}^{\infty }$,$\left\{ {{b}_{n}}
\right\}_{n=1}^{\infty }$ be the sequences generated by alternating
projection onto $A$ and $B$ from any intimal point ${{x}_{0}}\in
{\mathbb{R}^{n}}$:
\begin{eqnarray}
{{a}_{n}} &=& {{P}_{A}}\left( {{b}_{n-1}} \right),\\
{{b}_{n}} &=& {{P}_{B}}\left( {{a}_{n}} \right),\\
{{a}_{1}} &=& {{P}_{A}}\left( {{x}_{0}} \right).
\end{eqnarray}
1. If $A\bigcap B\ne \O $,\begin{equation} {{a}_{n}},{{b}_{n}}\to
{{x}^{*}}\in A\bigcap B.
\end{equation}
2. if $A\bigcap B=\O$,\begin{equation} {{a}_{n}}\to {{a}^{*}}\in
A,{{b}_{n}}\to {{b}^{*}}\in B,
\end{equation}
where $\left\| {{a}^{*}}-{{b}^{*}} \right\|=dist(A,B)$.
\end{thm}

Ordinary alternating projection can only achieve some point
arbitrarily on the intersection but not the orthogonal projection,
so we employ another variant projection algorithm$-$Dykstra's
alternating projection. This method is usually employed to the
problem

\begin{equation} {\underset{x\in
R}{\text{minimize}}}\,{{\left\| x-r \right\|}^{2}}\ subject\ to\
x\in \bigcap _{i=1}^{n}{{A}_{i}},
\end{equation}
which provides the best approximations to the sets.

Recently, Dykstra's algorithm has been extended to solve
least-squares \cite{escalante1996dykstra}, convex optimization
\cite{boyd2011distributed}, etc..

Dykstra's alternating projection implements correction at each
projection to Bregman's method by subtracting the variable, i.e.,
increment. Following theorem provides detailed descriptions.

\begin{thm}\textsuperscript{\cite{cao2012overview}}
Let ${{A}_{1}},{{A}_{2}},\ldots {{A}_{n}}\subseteq {{R}^{n}}$ be the
closed convex sets with nonempty intersection. Given $x\in {R}^{n}$
iterate by
\begin{eqnarray}
x_{n}^{i}&:=&{{P}_{{{A}_{i}}}}\left( x_{n}^{i-1}-I_{n-1}^{i}\right),\\
I_{n}^{i}&:=&x_{n}^{i}-\left( x_{n}^{i-1}-I_{n-1}^{i} \right),\\
x_{n}^{0}&:=&x_{n-1}^{r},
\end{eqnarray}
with initial values $x_{1}^{0}:=x$, $I_{0}^{i}:=0$ then

\begin{equation}
{{x}_{n}}\to {{P}_{\bigcap _{i=1}^{n}{{A}_{i}}}}\left( x \right).
\end{equation}

\end{thm}

\section{Algorithm}
In this section, we will firstly transform the minimum time
rendezvous problem into searching for the distance between the
zero-time plane and the intersection of several second-order cones.
Secondly, the problem is handled with a 2-step alternating method in
distributed setting.
\subsection{Problem based on geometric description}

The minimum time cost by vehicle $i$ from the initial position
$\left({{x}_{i}},{{y}_{i}}\right)$ with speed of $v$ to any specific
position $p:\left( x,y \right)$ in the 2-D plane is
\begin{equation}
\label{MinTimeToPoint} {{t}_{i}}\left( x,y \right)=\frac{{{\left\|
\left( x,y \right)-\left( {{x}_{i}},{{y}_{i}} \right)
\right\|}_{2}}}{{{v}_{i}}}.
\end{equation}

Note that in the  minimum time rendezvous problem
\eqref{MTRPhetervel}, ${\mathop {\max }\limits_i \frac{{\left\|
{\left( {x,y} \right) - \left( {{x_i},{y_i}} \right)}
\right\|}}{{{v_i}}}}$ is the point-wise maximum to
\eqref{MinTimeToPoint}. From \cite{boyd2004convex}, we know that
because \eqref{MinTimeToPoint} is convex in position, the epigraph
of the point-wise maximum corresponds to the intersection of
epigraphs of \eqref{MinTimeToPoint}:

\begin{equation}
\mathbf{epi}\left\{ \underset{i}{\mathop{\max }}\,\left(
{{t}_{i}}\left( x,y \right) \right)
\right\}=\underset{i}{\mathop{\cap }}\,\mathbf{epi}\left\{
{{t}_{i}}\left( x,y \right) \right\}
\label{MaxEpiToIntersectionEpi}.
\end{equation}

Since the potential arrival time $t$ to any position must be equal
to or larger than the minimum time, the epigraph of
\eqref{MinTimeToPoint} is the potential arrival time. The potential
time for vehicle $i$ to any positions forms a second-order cone in
position-time space,
\begin{equation}
{{C}_{i}}:t=\mathbf{epi}\left\{ {{t}_{i}}\left( x,y \right)
\right\}\ge \frac{{{\left\| \left( x,y \right)-\left(
{{x}_{i}},{{y}_{i}} \right) \right\|}_{2}}}{{{v}_{i}}},
\label{epiCone}
\end{equation}
with the initial positions as apexes and ${v}_{i}$ as the slope of
the generatrix.
%


Applying \eqref{MaxEpiToIntersectionEpi} and \eqref{epiCone} to the
original problem \eqref{MTRPhetervel}, we can transform
\eqref{MTRPhetervel} into,
\begin{equation}
\underset{x,y}{\mathop{\min }}\,\underset{i}{\mathop{\bigcap
}}\,{{C}_{i}}. \label{MinInterCone}
\end{equation}
Therefore, the minimum time rendezvous point becomes the lowest
point of the intersection of the cones. In other words, the minimum
time is the distance between the intersection and the zero-time
plane in the position-time space.


\subsection{2-step seeking method}

From the discussions in Section II, we know that Bregman's
alternating algorithm is employed to achieve the distance between
two disjoint convex sets. Beside zero-time plane, the intersection
in \eqref{MinInterCone} is also a convex set, because the
intersection of a finite number of second-order cones is convex.
Furthermore, rendezvous time for vehicles starting from different
points must be larger than zero, and consequently there is no common
point in the zero-time plane and the intersection. Therefore, the
Bregman's alternating algorithm is absolutely applicable here to
find the distance. Nevertheless, during this procedure, Bregman's
algorithm requires the orthogonal projection from a given point onto
the intersection, which is difficult to obtain especially in
distributed setting. Dykstra's alternating projection in Theorem 2
provides an efficiency method to such problem. As mentioned before,
we are about to engage Dykstra's algorithm as the intermediate
procedure in Bregman's algorithm.With cyclic digraph interaction
topology, the vehicles utilize Dykstra's algorithm to reach
consistency on the projection onto the interaction and Bregman's
algorithm to ensure consensus on the minimum time rendezvous point.

%
%

The algorithm is detailed in the following table.

\begin{table}[h]
\centering
\renewcommand{\arraystretch}{1.2}
\renewcommand{\thefootnote}{\alph{footnote}}
\begin{minipage} {0.5\textwidth}
\begin{tabular}{p{3.2in}} \hline
\textbf{Algorithm}: 2-steps alternating projection
\\ \hline
${{x}_{t0}}={{x}_{init}}(i)$\ \{Initial guess from vehicle's start
position\} \\
${{y}_{t0}}={{y}_{init}}(i)$\\
${{t}_{t0}}=0$\\
${{I}_{t0}}=\left[
\begin{matrix}
   0 & 0 & 0  \\
\end{matrix} \right]$\ \{Initial increment\}\\
\textbf{For} $l=1$ \textbf{To} $MaxIteration$ \textbf{Do} \\
\quad \textbf{If} vehicle number $i\ne 1$ \textbf{Then}\ \{If not the first vehicle\}\\
\quad\quad $\left( {{x}_{pre}},{{y}_{pre}},{{t}_{pre}}
\right)=ReceivePrevious\left( i-1
      \right)$\\
\quad\quad \{Vehicle 1 receive information from the last one\}\\
\quad \textbf{Else}\\
\quad\quad $\left( {{x}_{pre}},{{y}_{pre}},{{t}_{pre}}
\right)=\text{ }ReceivePrevious\left( LastVehicle
      \right)$\\
\quad\quad \{Receive information from previous vehicle\}\\
\quad \textbf{End If}\\
\quad $\left( {{x}_{0}},{{y}_{0}},{{t}_{0}} \right)=cone\_projection$ ((${{x}_{pre}},{{y}_{pre}},{{t}_{pre}}$)$-{{I}_{tl-1}},$\\
\hspace{16em} ${{x}_{t0}},{{y}_{t0}},v$)\\
\quad ${{I}_{tl}}=\left( {{x}_{tl}},{{y}_{tl}},{{t}_{tl}}
\right)-\left[ \left( {{x}_{pre}},{{y}_{pre}},{{t}_{pre}}
\right)-{{I}_{tl-1}}
    \right]$\\
\quad \{Dykstra's alternating projection method\}\\
\quad \textbf{If} $mod\left( l,BregmanFrequence \right)=0$ \textbf{Then}\\
\quad \{Time for proceeding Bregman's alternating projection method\}\\
\quad\quad ${{I}_{tl}}=\left( 0,0,0 \right)$\ \{Reset increment and time to zero\}\\
\quad\quad ${{t}_{tl}}=0$\\
\quad\quad \textbf{If} $i\ne 1$ \textbf{Then}\ \{Vehicle 1 conduct Bregman's alternating\\
\hspace{8em} projection to zero time plane\}\\
\quad\quad\quad ${{x}_{tl}}=0$\\
\quad\quad\quad ${{y}_{tl}}=0$\\
\quad\quad \textbf{End If}\\
\quad \textbf{End If}\\
\quad $SendtoNext \left( {{x}_{tl}},{{y}_{tl}},{{t}_{tl}} \right)$\ \{Send the projection point to next\\
\hspace{12.5em} vehicle\}\\
\textbf{End For}\\ \hline

\end{tabular}
\end{minipage}
\end{table}

The function 'cone\_projection' in the algorithm is used to obtain
the projection onto a cone.

\begin{table}[H]
\centering
\renewcommand{\arraystretch}{1.2}
\renewcommand{\thefootnote}{\alph{footnote}}
\begin{minipage} {0.5\textwidth}
\begin{tabular}{p{3.2in}} \hline
\textbf{Cone\_Projection Function}: $\left({{x}_{0}},{{y}_{0}},{{t}_{0}} \right)=cone\_projection$ \\
\hspace{12em} $\left( \left(
{{x}_{pre}},{{y}_{pre}},{{t}_{pre}}\right)-{{I}_{tl-1}},{{x}_{t0}},{{y}_{t0}},v
\right)$\\ \hline

\textbf{Input}:Target Position $p$ and time $t$ for projection,the cone's apex $p0$,\\
\qquad\quad the slope of the generatrix (vehicle's velocity) $v$ \\

\quad $s=norm\left( p-p0 \right)$ \ \{Euclidean distance between original position\\
\hspace{10em} to apex\} \\

\quad \textbf{If} $\left( \left( s-abs\left( v*t \right) \right)>0 \right)$ \textbf{Then}\ \{Projection onto the surface of the\\
\hspace{14.5em} cone\}\\
\quad \quad ${{p}_{p}}=(p-p0)\frac{({{v}^{2}}s+vt)}{({{v}^{2}}s+s)}+p0$\\
\quad \quad ${{t}_{p}}=\frac{(vs+t)}{({{v}^{2}}+1)}$\\
\quad \textbf{Else If} $\left( s<=vt \right)$ \textbf{Then}\ \{Original point inside the cone\}\\
\quad \quad $p\_p=p$\ \{Projection the same as original\}\\
\quad \quad $t\_p=t$\ \\
\quad \textbf{Else If} $\left( s<=-vt \right)$ \textbf{Then}\ \{Projection onto apex\}\\
\quad \quad $p\_p=p0$\\
\quad \quad $t\_p=0$\\
\quad \textbf{End If} \\
\quad $\left( {{x}_{pre}},{{y}_{pre}} \right)=p\_p$\ \{Distribute position to components\}\\
\quad ${{t}_{pre}}=t\_p$\\
\quad \textbf{Return} $\left({{x}_{pre}},{{y}_{pre}},{{t}_{pre}} \right)$\\
\hline
\end{tabular}
\end{minipage}
\end{table}

\begin{rem}
The frequency of Bregman's alternating projection applied to the
vehicles ensures the converge speed, and the frequency of Dykstra's
alternating projection guarantees the accuracy.
\end{rem}

\begin{rem}
When the items 'BregmanFrequence' in Algorithm 1 are set differently
for the vehicles, vehicles execute in an asynchronous setting.
Although this configuration may seem to damage the cooperation,
algorithms still function well, and the results are given in the
next section.
\end{rem}

\section{Simulation}

To illustrate efficiency of our algorithm, we provide following
demonstrations. Consider five vehicles, whose initial positions and
velocities are $\left( 0m,0m \right)$, $\left( 100m,20m \right)$,
$\left( 150m,200m \right)$, $\left( 50m,50m \right)$, $\left(
20m,170m \right)$, and $5m/s$, $7m/s$, $10m/s$, $6m/s$, $4m/s$,
respectively. The minimum time rendezvous point is
$\left(50m,66m\right)$ and the time is $16.6667s$.

We apply our algorithm to those 5 vehicles in following two
scenarios. In the first scenario, the five vehicles know how many
iterations have been proceeded, and process the Bregman's projection
(i.e., make the increments reset to zero) simultaneously. The second
scenario studies the asynchronous case, where the agents have their
own clock to reset the increments.


\subsection{Scenario I: Simultaneous Bregman's projecting}

\begin{figure}[H]
\centering
\includegraphics[width=3in]{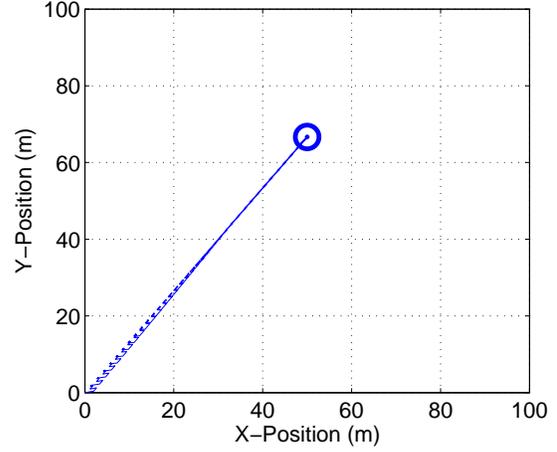}
\caption{ Trajectory of the agreements to the rendezvous point in
x-y plane. } \label{track}
\end{figure}

\begin{figure}[H]
\centering
\includegraphics[width=3in]{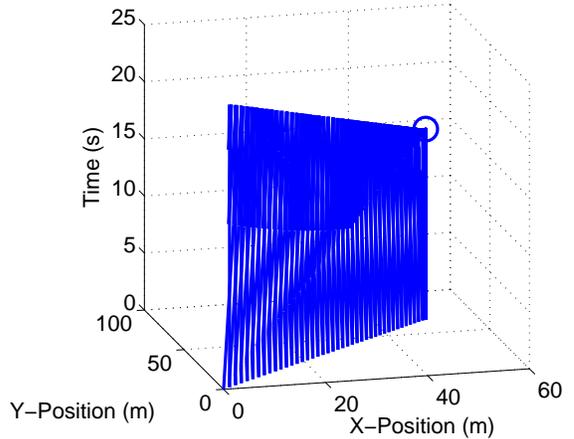}
\caption{ Trajectory of the agreements to the rendezvous with time.
} \label{track3d}
\end{figure}

Fig.\ref{track} shows the position evolution history during the
interactions between the vehicles. Fig.\ref{track3d} presents
projection points in position-time space. Five vehicles can all
obtain the minimum time rendezvous point, the circle in the figures.
As shown in the Fig.\ref{track3d}, the line oscillates, because the
Bregman's alternating projection switches between the x-y-0 plane
and the intersection of the cones, and finally the oscillate between
the two points yield the distance, or the minimum time.


\begin{figure}[H]
\centering
\includegraphics[width=3in]{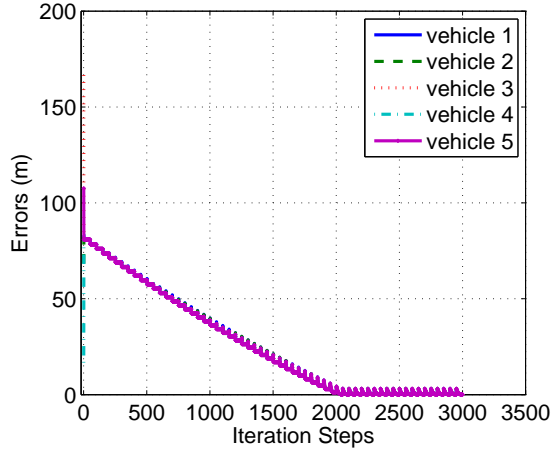}
\caption{ Error to the real minimum time rendezvous point with low
frequency Bregman's projection. } \label{errors}
\end{figure}

\begin{figure}[H]
\centering
\includegraphics[width=3in]{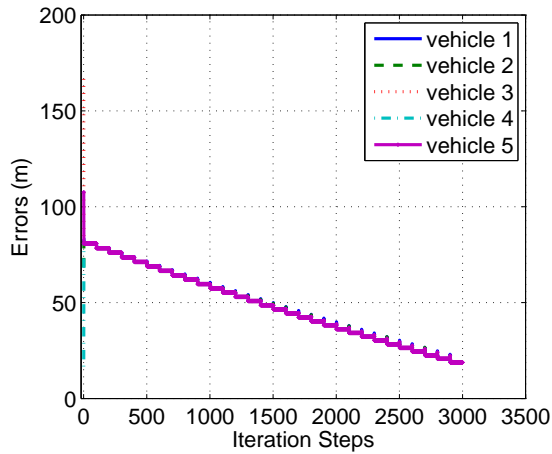}
\caption{ Error to the real minimum time rendezvous point with high
frequency Bregman's projection. } \label{error_slow}
\end{figure}

Fig.\ref{errors} and Fig.\ref{error_slow} present the errors to the
actual minimum time rendezvous point, $\left(50m,66m\right)$, during
the procedure. In Fig.\ref{errors} the Bregman's projection is
applied every time after the Dykstra's projection proceeds 50
cycles, and in Fig.\ref{error_slow} 100 cycles. As the results
shown, the frequency of Bregman's projection applied not only
affects the convergent rate but also the accuracy. This is due to
the fact that reducing the frequency of Bregman's projection means
more Dykstra's projection would be used, and more accuracy of the
projection onto the intersection of the cones would be achieved.
Conversely, Bregman's projection will quickly approach the solution
with a rough estimate.


\subsection{Scenario II: Asynchronous Bregman's projecting}

In this scenario, we let vehicle 1 reset its increment to zero every
50 cycles, vehicle 2 and 3 every 40 cycles, and vehicle 4 and 5
every 75 cycles. This setting will result in error on increments in
the procedure of Dykstra's projection.

\begin{figure}[H]
\centering
\includegraphics[width=3in]{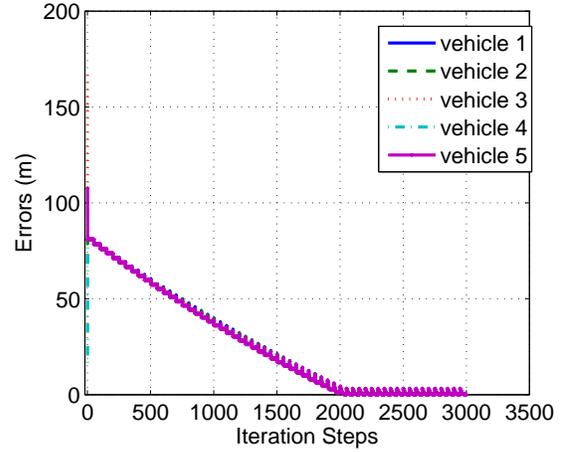}
\caption{ Error to the real minimum time rendezvous point with
asynchronous Bregman’s projecting. } \label{Asy_Error}
\end{figure}

\begin{figure}[H]
\centering
\includegraphics[width=3 in]{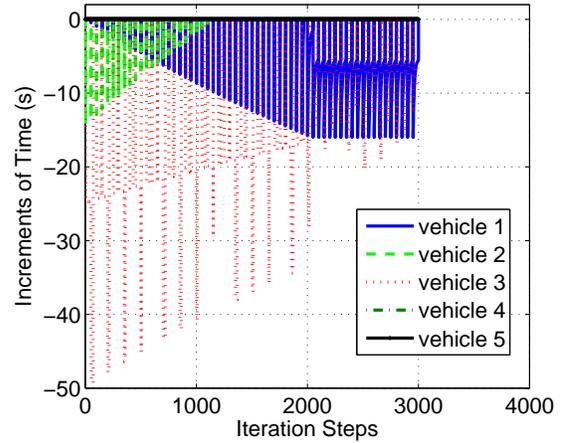}
\caption{ Increments of each vehicle when asynchronous Bregman’s
projecting applied. } \label{Asy_Inc}
\end{figure}

As shown in Fig.\ref{Asy_Error}, the errors also converge to zero
with around 2000 times interactions, about the same as the
simultaneous case. Fig.\ref{Asy_Inc} illustrates the increment of
time changes when the algorithm is applied. Although different
frequency of reset is employed, the increments perform as if a
global synchronous clock triggers Bregman's projection. We speculate
this as the result of nonlinear interactions within the vehicle.

\section{CONCLUSIONS AND FUTURE WORKS}

We have proposed the distributed algorithm on multi-vehicle minimum
time rendezvous point seeking which is composed of the Bregman's and
Dykstra's alternating projection method. We have shown that the
distance between the position plane and the intersection of the
second-order cones is the minimum time for rendezvous, and the point
on the plane that achieves the distance is the minimum time
rendezvous point. The Bregman's alternating projection was used to
obtain the distance by consistent projections onto the plane and the
intersection of the cones. Dykstra's method guaranteed the vehicles
with
cyclic digraph interaction network had consensus on the projection
of the point onto the intersection of the cones and only the
communication with the neighbor is applied. The frequency of
Bregman's alternating projection applied to the vehicles ensures the
converge speed, and the frequency of Dykstra's alternating
projection ensures the accuracy. The increments should be reset to
zero after the procedure of cyclic Dykstra's method, but the
demonstration result showed that the vehicles could make the
prediction on when to apply Bregman's projection and reset the
increment asynchronously. We believe the ability of asynchronous is
linked to the characteristics of the alternating projection. This
leaves the question of robustness of asynchronous open. In future
research, we will analyze asynchronous character, and consider the
interaction network with loose constraints.

\section{ACKNOWLEDGMENTS}



\begin{thebibliography}{99}

\bibitem{cao2012overview}
Y.~Cao, W.~Yu, W.~Ren, and G.~Chen, ``An overview of recent progress
in the
  study of distributed multi-agent coordination,'' 2012.

\bibitem{ren2005survey}
W.~Ren, R.~W. Beard, and E.~M. Atkins, ``A survey of consensus
problems in
  multi-agent coordination,'' in \emph{American Control Conference, 2005.
  Proceedings of the 2005}.\hskip 1em plus 0.5em minus 0.4em\relax IEEE, 2005,
  pp. 1859--1864.

\bibitem{kranakis2006mobile}
E.~Kranakis, D.~Krizanc, and S.~Rajsbaum, ``Mobile agent rendezvous:
A
  survey,'' in \emph{Structural Information and Communication
  Complexity}.\hskip 1em plus 0.5em minus 0.4em\relax Springer, 2006, pp. 1--9.

\bibitem{mclain2001cooperative}
T.~W. McLain, P.~R. Chandler, S.~Rasmussen, and M.~Pachter,
``Cooperative
  control of uav rendezvous,'' in \emph{American Control Conference, 2001.
  Proceedings of the 2001}, vol.~3.\hskip 1em plus 0.5em minus 0.4em\relax
  IEEE, 2001, pp. 2309--2314.

\bibitem{lin2003multi}
J.~Lin, A.~S. Morse, and B.~D. Anderson, ``The multi-agent
rendezvous
  problem,'' in \emph{Decision and Control, 2003. Proceedings. 42nd IEEE
  Conference on}, vol.~2.\hskip 1em plus 0.5em minus 0.4em\relax IEEE, 2003,
  pp. 1508--1513.

\bibitem{lin2004multi}
J.~Lin, A.~S. Morse, and B.~D. Anderson, ``The multi-agent
rendezvous problem-the asynchronous case,'' in
  \emph{Decision and Control, 2004. CDC. 43rd IEEE Conference on},
  vol.~2.\hskip 1em plus 0.5em minus 0.4em\relax IEEE, 2004, pp. 1926--1931.

\bibitem{cortes2006robust}
J.~Cort{\'e}s, S.~Mart{\'\i}nez, and F.~Bullo, ``Robust rendezvous
for mobile
  autonomous agents via proximity graphs in arbitrary dimensions,''
  \emph{Automatic Control, IEEE Transactions on}, vol.~51, no.~8, pp.
  1289--1298, 2006.

\bibitem{fang2008multi}
J.~Fang, A.~S. Morse, and M.~Cao, ``Multi-agent rendezvousing with a
finite set
  of candidate rendezvous points,'' in \emph{American Control Conference,
  2008}.\hskip 1em plus 0.5em minus 0.4em\relax IEEE, 2008, pp. 765--770.

\bibitem{hui2011finite}
Q.~Hui, ``Finite-time rendezvous algorithms for mobile autonomous
agents,''
  \emph{Automatic Control, IEEE Transactions on}, vol.~56, no.~1, pp. 207--211,
  2011.

\bibitem{meschler1963time}
P.~Meschler, ``Time-optimal rendezvous strategies,'' \emph{Automatic
Control,
  IEEE Transactions on}, vol.~8, no.~4, pp. 279--283, 1963.

\bibitem{chyung1973time}
D.~Chyung, ``Time optimal rendezvous of three linear systems,''
\emph{Journal
  of Optimization Theory and Applications}, vol.~12, no.~3, pp. 242--247, 1973.

\bibitem{brown2011determination}
T.~L. Brown, T.~D. Aslam, and J.~P. Schmiedeler, ``Determination of
minimum
  time rendezvous points for multiple mobile robots via level set methods,'' in
  \emph{ASME 2011 International Design Engineering Technical Conferences and
  Computers and Information in Engineering Conference}.\hskip 1em plus 0.5em
  minus 0.4em\relax American Society of Mechanical Engineers, 2011, pp.
  787--797.

\bibitem{bhatia2008decentralized}
A.~Bhatia and E.~Frazzoli, ``Decentralized algorithm for
minimum-time
  rendezvous of dubins vehicles,'' in \emph{American Control Conference,
  2008}.\hskip 1em plus 0.5em minus 0.4em\relax IEEE, 2008, pp. 1343--1349.

\bibitem{crispin2008interception}
Y.~J. Crispin, ``Interception and rendezvous between autonomous
vehicles,''
  \emph{Advances in Robotics, Automation and Control Vienna, Austria: InTech
  Publishing KG}, 2008.

\bibitem{luo2006optimization}
Y.-Z. Luo, G.-J. Tang, and H.-y. Li, ``Optimization of
multiple-impulse
  minimum-time rendezvous with impulse constraints using a hybrid genetic
  algorithm,'' \emph{Aerospace science and technology}, vol.~10, no.~6, pp.
  534--540, 2006.

\bibitem{paiewonsky1966three}
B.~Paiewonsky and P.~J. Woodrow, ``Three-dimensional time-optimal
rendezvous.''
  \emph{Journal of Spacecraft and Rockets}, vol.~3, no.~11, pp. 1577--1584,
  1966.

\bibitem{burns2007autonomous}
B.~S. Burns, P.~A. Blue, and M.~D. Zollars, ``Autonomous control for
automated
  aerial refueling with minimum-time rendezvous,'' in \emph{Proceedings of AIAA
  Guidance, Navigation and Control Conference}, 2007.

\bibitem{wilson2013real}
D.~B. Wilson, M.~A.~T. Soto, A.~H. Goktogan, and S.~Sukkarieh,
``Real-time
  rendezvous point selection for a nonholonomic vehicle,'' in \emph{Robotics
  and Automation (ICRA), 2013 IEEE International Conference on}.\hskip 1em plus
  0.5em minus 0.4em\relax IEEE, 2013, pp. 3941--3946.

\bibitem{notarstefano2006distributed}
G.~Notarstefano and F.~Bullo, ``Distributed consensus on enclosing
shapes and
  minimum time rendezvous,'' in \emph{Decision and Control, 2006 45th IEEE
  Conference on}.\hskip 1em plus 0.5em minus 0.4em\relax IEEE, 2006, pp.
  4295--4300.

\bibitem{tam2012method}
M.~K. Tam, ``The method of alternating projections,'' Ph.D.
dissertation,
  University of Newcastle, Australia, 2012.

\bibitem{combettes2011proximal}
P.~L. Combettes and J.-C. Pesquet, ``Proximal splitting methods in
signal
  processing,'' in \emph{Fixed-point algorithms for inverse problems in science
  and engineering}.\hskip 1em plus 0.5em minus 0.4em\relax Springer, 2011, pp.
  185--212.

\bibitem{boyd2003alternating}
S.~Boyd and J.~Dattorro, ``Alternating projections,'' \emph{Lecture
notes of EE
  392 o, Stanford University, Autumn Quarter}, vol. 2004, 2003.

\bibitem{escalante1996dykstra}
R.~Escalante and M.~Raydan, ``Dykstra's algorithm for a constrained
  least-squares matrix problem,'' \emph{Numerical linear algebra with
  applications}, vol.~3, no.~6, pp. 459--471, 1996.

\bibitem{boyd2011distributed}
S.~Boyd, N.~Parikh, E.~Chu, B.~Peleato, and J.~Eckstein,
``Distributed
  optimization and statistical learning via the alternating direction method of
  multipliers,'' \emph{Foundations and Trends{\textregistered} in Machine
  Learning}, vol.~3, no.~1, pp. 1--122, 2011.

\bibitem{boyd2004convex}
S.~P. Boyd and L.~Vandenberghe, \emph{Convex optimization}.\hskip
1em plus
  0.5em minus 0.4em\relax Cambridge university press, 2004.

\end{thebibliography}
\end{document}